\documentclass[prl, floatfix, twocolumn, nofootinbib]{revtex4}
\usepackage{graphicx}
\usepackage{color}
\usepackage{amsmath, amsfonts, amssymb, bm}
\usepackage{gensymb}
\usepackage{braket}
\usepackage{subfigure}
\begin{document}
\title{
Interatomic distance dependence of resonant energy-transfer phenomena} 
\author{F. Gr\"ull}
\author{A. B. Voitkiv}
\author{C. M\"uller}
\affiliation{Institut f\"ur Theoretische Physik I, Heinrich Heine Universit\"at D\"usseldorf, 
Universit\"atsstr. 1, 40225 D\"usseldorf, Germany}
\date{\today}
\begin{abstract}
It is well known that interatomic or intermolecular interactions driven by 
two-center electronic dipole-dipole correlations fall off rapidly with the 
inter-site distance. We show, however, that the effective strength of 
interatomic reaction channels, which are triggered by a resonant field, 
can exhibit a nonmonotonous distance dependence, being strongly reduced when 
the atoms come closer. This surprising result is demonstrated by considering 
resonant two-center photoionization as an example. Our findings are supported 
by available experimental data.
\end{abstract}
\maketitle

{\it Introduction---}Interatomic and intermolecular processes are under 
very active scrutiny in recent years. The research area has been strongly 
triggered by the prediction of interatomic Coulombic decay (ICD) where 
electronic excitation energy of an atom is transferred radiationlessly
to a neighbor atom resulting in its ionization \cite{ICD, ICDres, ICDrev}. 
ICD is of particular importance, when a single-center Auger decay is 
energetically forbidden, and can proceed much faster than radiative decay.
It has first been observed in noble gas dimers and clusters \cite{clusters, dimers}. 
Corresponding measurements rely on advanced experimental techniques, such as 
third-generation synchrotron sources and few-body coincidence spectrometers \cite{COLTRIMS}.

Similar interatomic energy-transfer processes are known in various areas 
of physics, comprising exciton dynamics in solids \cite{excitons}, 
quantum optical ensembles and cold Rydberg gases \cite{Rydberg}. They also 
play an important role in chemistry and biology, as exemplified by ICD in 
water \cite{water} and hydrated biomolecules \cite{Dorn}, lattice dynamics 
in polymers \cite{polymers}, 
and F\"orster resonances in chromophores \cite{Forster}. Slow electrons
set free via ICD cascades are of great relevance for applied radiation 
biology \cite{Gokhberg}. It has thus been concluded that interatomic 
energy-transfer reactions are ubiquitous in nature.

In most of the cases, the interatomic coupling arises from the long-range
interaction between two dipoles. It is of the form (in atomic units)
\begin{eqnarray} 
\hat{V}_{ee} = \frac{{\bf r}\cdot \boldsymbol{\xi}}{R^3} 
- \frac{ 3 ({\bf r}\cdot{\bf R})(\boldsymbol{\xi}\cdot{\bf R})}{R^5}\ ,
\label{V_AB} 
\end{eqnarray} 
with the internuclear separation ${\bf R}$ and the coordinates of the active electrons 
${\bf r}$ and ${\bf r}'={\bf R}+\boldsymbol{\xi}$. The generic $R^{-3}$ scaling is 
modified when retardation effects or non-dipole transitions are considered. Yet always 
the interaction quickly falls off when the interatomic distance grows. Accordingly, 
inter-site energy transfers are expected to be the more efficient, the closer the
atoms lie together.

An interatomic process involving ICD is two-center resonant photoionization (2CPI) 
in a heteroatomic system of two atoms, say $A$ and $B$ 
\cite{2CPI, 2CPI_PRA, Perina, 2CPIcoll, 2CPImol}. Here, a neighboring atom $B$ is 
first resonantly excited by photoabsorption, this way creating an autoionizing state 
of the two-center system, which afterwards stabilizes via ICD. The original 
theory \cite{2CPI, 2CPI_PRA} assumed, for simplicity, two spatially well-separated atoms 
with fixed distance vector ${\bf R}$. Since the photoabsorption step is included in the 
treatment, a comparison with the direct photoionization of atom $A$ is feasible. 
Application to Li as atom $A$ and He as atom $B$ at an 
interatomic distance of 10\,\AA\ as an example, a relative enhancement of 
2CPI over the direct photoionization of Li by a factor $\simeq 10^6$ is predicted.

2CPI was experimentally observed in NeHe dimers \cite{2CPIexp, Jabbari, Mhamdi2018}.
Up to a $\simeq$\,100-fold   
relative enhancement was found, which is very substantial, though 
much smaller than the enormous amplification predicted for the LiHe model system. This 
discrepancy appears astonishing in light of the fact that the internuclear separation 
in the NeHe ground state is substantially less than in LiHe lying between 
$\approx 2.5$--6\,\AA\ \cite{Sisourat2010}. Accordingly, 
ICD proceeds much faster in NeHe than in LiHe, occuring on a timescale of hundreds of femtoseconds. 
A considerable enhancement of photoionization due to 2CPI has very recently also been seen 
in ArNe clusters \cite{Hergenhahn}, but again at much lower scale than in LiHe.

The reasons for so vastly different (and counterintuitive) levels of enhancement 
have not been clarified yet, nor has a theoretical explanation for the observed 
levels been given. 
It is suggestive to assume that the differences between the measurements and the original 
theory result from the molecular structure of the target systems which was not taken into 
account there. In fact, ICD in NeHe and other noble-gas dimers is known to be very sensitive 
to the vibrational nuclear motion \cite{Sisourat2010, He-dimer}.

However, as we show in this paper, the difference between the relative enhancement 
predicted in LiHe versus the one observed in NeHe is not caused by the molecular 
structure of a dimer in the first place. The reduction in 2CPI efficiency rather occurs 
because the Ne and He atoms are {\it so close} to each other.
This result stands in sharp contrast to the intuitive expectation that close distances
should generally facilitate interatomic energy transfer processes [see Eq.~\eqref{V_AB}]. 
Nevertheless, it can be obtained within a relatively simple theoretical treatment of 
the 2CPI process in a weakly bound dimer, which explains why, close to the resonance,  
a large ICD rate can have a detrimental impact on the effectiveness of 2CPI.

\begin{figure}[b]  
\vspace{-0.25cm}
\begin{center}
\includegraphics[width=0.4\textwidth]{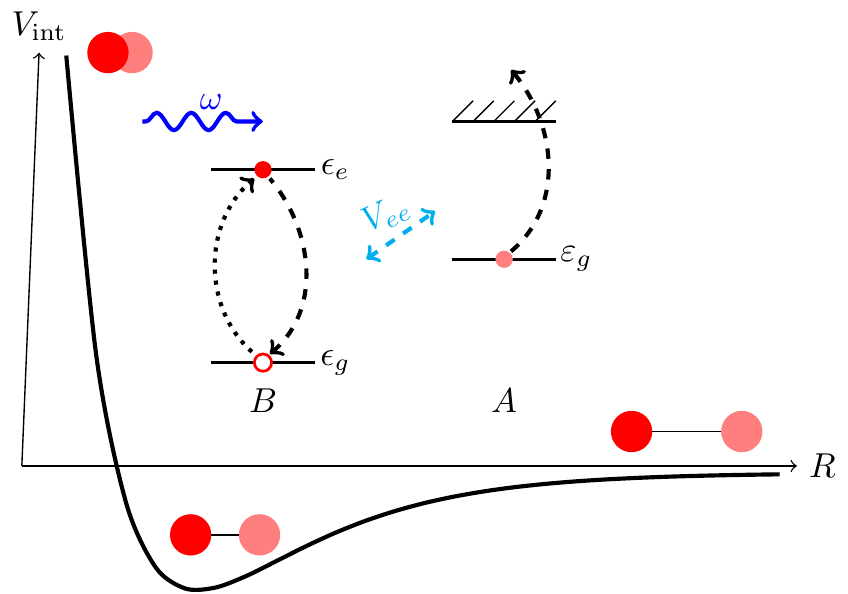}
\end{center}
\vspace{-0.5cm} 
\caption{ Scheme of two-center resonant photoionization (2CPI), embedded in a generic
potential curve $V_{\rm int}(R)$ of a van-der-Waals dimer $A$-$B$. First, atom $B$ is 
resonantly photoexcited; its subsequent decay by ICD leads to ionization of atom $A$ 
via an interatomic dipole-dipole interaction $V_{ee}$. The shape of the potential 
curve determines at which internuclear distance $ R $ the process mostly occurs. }
\label{figure1}
\end{figure}

{\it Theoretical description---}We consider a system of two atoms, $A$ and $B$, 
which are initially in their ground states. They are separated by a sufficiently 
large distance $R$ (covering at least several \AA), such that their individuality 
is basically preserved, and exposed to a resonant electromagnetic field of the form
\begin{eqnarray}
{\bf F}(t)= F_0 \cos(\omega t)\,{\bf e}_z
\label{field}
\end{eqnarray}
which is taken in the dipole approximation. Here, $\omega$ is the angular frequency 
and $F_0$ the field amplitude.

To start with, we assume the atoms to be at rest and 
take the position of the nucleus of atom $A$  
as the origin and denote the coordinates 
of the nucleus of atom $B$, the (active) electron of atom $A$ 
and that of atom $B$ by ${\bf R}$, ${\bf r}$
and ${\bf r}' ={\bf R} + \boldsymbol{\xi}$, 
respectively, where $\boldsymbol{\xi}$ 
is the position of the electron of atom $B$ 
with respect to its nucleus. 
Let atom $B$ have an excited state $\chi_e$
reachable from the ground state $\chi_g$ by a dipole-allowed transition. 

The total Hamiltonian describing the
two atoms in the external electromagnetic field reads 
\begin{eqnarray} 
\hat{H} =  \hat{H}_0 + \hat{V}_{ee} + \hat{W},  
\label{hamiltonian}  
\end{eqnarray} 
where $ \hat{H}_0 $ is the sum of the Hamiltonians 
for the noninteracting atoms $A$ and $B$, and
$\hat{V}_{ee}$ the interaction between the atoms. 
$\hat{W} = \hat{W}_A + \hat{W}_B = {\bf F}(t) \cdot ({\bf r}+{\bf r}')$ 
denotes the interaction of the atoms with the electromagnetic field
in the length gauge. It is assumed that $\omega_{ge} R /c  \ll 1$, 
where $\omega_{ge}$ is the atomic transition frequency and $c$ the 
speed of light, such that retardation effects can be neglected.

In the 2CPI process one has essentially three different basic two-electron 
configurations, which are schematically illustrated in Fig.~\ref{figure1}:
(I) $\Phi_i = \varphi_{g}({\bf r}) \chi_g(\boldsymbol{\xi})$ with total energy 
$E_i = \varepsilon_g + \epsilon_g$, where both atoms are in the corresponding 
ground states $\varphi_g$ and $\chi_g$;
(II) $\Phi_a = \varphi_{g}({\bf r}) \chi_e(\boldsymbol{\xi})$ with total energy 
$E_a = \varepsilon_g + \epsilon_e$, in which atom $A$ is in the ground state 
while atom $B$ is in the excited state $\chi_e$;
(III) $\Phi_f = \varphi_{\bf p}({\bf r}) \chi_g(\boldsymbol{\xi})$ with total energy 
$E_f = \varepsilon_{p} + \epsilon_g$, where the electron of atom $A$ has been 
emitted into the continuum with asymptotic momentum ${\bf p}$, while the electron of atom $B$ 
has returned to the ground state. 

Within the second order of time-dependent perturbation theory, 
the probability amplitude for 2CPI can be written as
\begin{eqnarray}
S^{(2)}_{\bf p}\! &=&\! -\int_{-\infty}^{\infty} dt\, \langle \Phi_f|\hat{V}_{ee}| \Phi_a \rangle\, e^{-i(E_a-E_f)t}\nonumber\\
& & \times \int_{-\infty}^{t} dt'\, \langle \Phi_a|\hat{W}_{B}| \Phi_i \rangle\, e^{-i(E_i-E_a)t'}\ .
\label{S1}
\end{eqnarray}

Performing the time integrations, we obtain
\begin{eqnarray} 
S^{(2)}_{\bf p} &=& -i\pi\, \langle \varphi_{\bf p}|{\bf r}| \varphi_g \rangle\cdot\left( {\bf e}_z - \frac{3R_z}{R^2}\,{\bf R}\right) \nonumber\\
& & \times\ \frac{F_0}{R^3}\, \frac{\left| \langle \chi_e|\xi_z| \chi_g \rangle \right|^2}
{\Delta+\frac{i}{2}\Gamma}\ \delta(\varepsilon_{p} - \varepsilon_0 - \omega)\ ,
\label{S3}
\end{eqnarray}
where the detuning from the resonance $\Delta = \epsilon_g+\omega-\epsilon_e$ has been introduced
and the total width $\Gamma=\Gamma_r+\Gamma_a$ of the excited state $\chi_e$ in atom $B$ inserted. 
It accounts for the finite lifetime of this state and consists of the radiative width $\Gamma_r$ 
and the two-center Auger (ICD) width $\Gamma_a$ \cite{Gamma}. 
The delta function in Eq.~\eqref{S3} displays the law of energy conservation in the process.

From the transition amplitude we can obtain the total ionization cross section in the usual way by taking 
the absolute square, integrating it over the photoelectron momentum, and dividing it by the interaction 
time $\tau$ and the incident flux $j=\frac{cF_0^2}{8\pi \omega}$, that is
\begin{eqnarray} 
\sigma^{(2)}_{\rm at}(R) &=& \frac{1}{j\tau}\int\frac{{\rm d}^3p}{(2\pi)^3}\,\left|S^{(2)}_{\bf p}\right|^2\nonumber\\ 
&=& \int {\rm d}\Omega_p\, \big| \langle \varphi_{\bf p}|{\bf r}| \varphi_g \rangle \cdot\left( {\bf e}_z - 3\cos\theta_R\,{\bf e}_R\right) \big|^2 \nonumber\\ 
& &  \times\ \frac{\omega\,p}{2\pi c R^6}\ \frac{\left| \langle \chi_e|\xi_z| \chi_g \rangle \right|^4}{\Delta^2+\frac{1}{4}\Gamma^2}\ .
\label{CS}
\end{eqnarray}
Here, the value of $p$ is fixed by the energy conservation and we have introduced 
the unit vector ${\bf e}_R={\bf R}/R$ along the internuclear separation and the angle 
$\theta_R$ between ${\bf R}$ and the field direction \cite{eff_pol}. Equation~\eqref{CS} 
can be rewritten using the cross section 
$\sigma^{(1)}_{A} = \int {\rm d}\Omega_p\, \big| \langle \varphi_{\bf p}|z| \varphi_g \rangle\big|^2 
\frac{\omega\,p}{2\pi c}$ for the direct photoionization of atom $A$ by the electromagnetic field.
For the special cases, when the separation vector ${\bf R}$ between the atoms $A$ and $B$ 
is oriented either along the field direction or perpendicular to it, we obtain
\begin{eqnarray} 
\sigma^{(2)}_{\rm at}(R) &=& \frac{\alpha}{R^6}\, \frac{\left|\langle \chi_e|\xi_z| \chi_g \rangle \right|^4}{\Delta^2+\frac{1}{4}\Gamma^2} \,\sigma^{(1)}_A \nonumber\\
&=& \left(\frac{3\,\alpha c^3}{4\, \omega_{ge}^3 R^3}\right)^{\!2} \frac{\Gamma_{r,ge}^2}{\Delta^2+\frac{1}{4}\Gamma^2} 
\ \sigma^{(1)}_{A}
\label{CS2}
\end{eqnarray}
where $\alpha=2$ for ${\bf R}\parallel {\bf F}_0$ and $\alpha=1$ for 
${\bf R}\perp {\bf F}_0$. In the second step, the dipole matrix element 
has been expressed by the corresponding radiative width 
$\Gamma_{r,ge}=\frac{4\omega_{ge}^3}{3c^3}\left| \langle \chi_e|\xi_z| \chi_g \rangle \right|^2$ with $\omega_{ge}=\epsilon_e-\epsilon_g$. The compact formula \eqref{CS2} applies to 
two individual atoms at a distance $R$, carrying a single active electron each. 

Below we will consider diatomic systems containing helium as atom $B$. In this case, 
the two equivalent electrons must be described by appropriately symmetrized wave functions 
and their interaction with the field by a two-particle extension of the 
operator $\hat{W}_B$. This leads to an additional factor of 4 in the 2CPI cross section 
\cite{2CPImol}. Taking this into account and assuming that the field is exactly resonant 
($\Delta=0$), the ratio of the 2CPI and direct photoionization cross sections becomes 
\begin{eqnarray} 
\frac{\sigma^{(2)}_{\rm at}(R)}{\sigma^{(1)}_{A}} &=& \left(\frac{3\alpha c^3}{\omega^3 R^3}\right)^{\!2}
\frac{\Gamma_{r,ge}^2}{(\Gamma_r+\Gamma_a)^2} \, . 
\label{ratio_at}
\end{eqnarray}
By expressing the Auger width $\Gamma_a$ according to   
\begin{eqnarray} 
\Gamma_a(R) = \frac{3}{2\pi}\,\frac{c^4}{\omega^4 R^6}\,\Gamma_{r,ge}\,\sigma^{(1)}_A  
\label{Gamma_a}
\end{eqnarray}
(see, e.g., \cite{e-impact}), Eq.\,\eqref{CS2} can be put in a form 
which enables one a better understanding of the interatomic distance dependence of the 2CPI: 
\begin{eqnarray} 
\sigma^{(2)}_{\rm at}(R) = \frac{\alpha^2}{2}\,\sigma^{(\rm exc)}_{B}\,\frac{\Gamma_a\Gamma_{r,ge}}{(\Gamma_r+\Gamma_a)^2} , 
\label{branching}
\end{eqnarray} 
where $\sigma^{(\rm exc)}_{B} = 3\pi c^2/\omega^2$ \cite{exc} is the cross section 
for resonant photoexcitation of atom $B$. 

Apart from a numerical prefactor of order unity, Eq.\,\eqref{branching} can be represented as 
a product of two terms, $\sigma^{(\rm exc)}_{B} \Gamma_{r,ge}/(\Gamma_r+\Gamma_a)$ 
and $\Gamma_a/(\Gamma_r+\Gamma_a)$. 

The first of them describes the photoexcitation step of 2CPI resulting in 
the creation of the intermediate state. 
Since $\Gamma_{r,ge}/(\Gamma_r+\Gamma_a) <  \Gamma_{r,ge}/\Gamma_r $ it is seen that 
compared to an isolated atom $B$ ($\Gamma_a = 0$) the probability for the resonant excitation 
in the $A$-$B$ system is reduced due to a broadening of the resonance caused by the presence 
of the additional deexcitation pathway via ICD. 

The second term, which is simply a branching ratio, determines the probability that afterwards 
the intermediate state decays via ICD. Unlike the first one, it increases with $\Gamma_a$ 
approaching $1$ at $ \Gamma_a  \gg \Gamma_r$.   

At $ \Gamma_{r,ge} \simeq \Gamma_r $ (which, in particular, 
holds for the systems studied below)
the optimal enhancement of 2CPI over direct photoionization 
is reached for $ \Gamma_a \approx \Gamma_r $, strongly decreasing both  
at $\Gamma_a \ll \Gamma_r$ and $\Gamma_a \gg \Gamma_r$.  
Since $\Gamma_a$ falls with $R$, this suggests a nonmonotonous behavior of 2CPI on the dimer size: 
there is an 'optimal' value, where the efficiency of 2CPI is maximal and from where it
decreases not only towards larger but also towards smaller sizes.

\vspace{0.15cm} 

We now include effects of the nuclear motion in a weakly bound molecule. The Coulomb, 
exchange and van-der-Waals interactions between the atoms $A$ and $B$ create a static 
potential $V_{\rm int}(R)$, 
whose form depends on the electron configuration and in which the atomic nuclei occupy 
discrete vibrational levels. The wave functions of the system are accordingly amended, 
$\Psi_{i,a,f} = \Phi_{i,a,f}({\bf r},\boldsymbol{\xi}) \psi_{i,a,f}^{(\nu)}(R)$, to include the 
internuclear coordinate. When the derivation given above is repeated with these molecular states, 
the ratio of cross sections adopts the modified form \cite{2CPImol}
\begin{eqnarray} 
\frac{\sigma^{(2)}_{\rm mol}}{\sigma^{(1)}_{\rm mol}} = \frac{\sigma^{(2)}_{\rm at}(R_{\rm eq})}{\sigma^{(1)}_{A}}
\left(\frac{R_{\rm eq}^3{\rm FC}_{i,a}\left\langle\psi_f^{(\nu_f)}\big|R^{-3}\big|\psi_a^{(\nu_a)}\right\rangle}{{\rm FC}_{i,f}}\right)^{\!2}
\label{ratio_mol}
\end{eqnarray}
for a given set of vibrational quantum numbers $\nu_i$, $\nu_a$ and $\nu_f$. Here, 
the Franck-Condon factors 
\begin{eqnarray}
{\rm FC}_{i,a} = \left\langle\psi_a^{(\nu_a)}\big|\psi_i^{(\nu_i)}\right\rangle 
= \int {\rm d}R\, [\psi_a^{(\nu_a)}(R)]^*\psi_i^{(\nu_i)}(R)
\label{FC}
\end{eqnarray}
and similarly for ${\rm FC}_{i,f}$ have been introduced. The quadratic factor 
in Eq.\,\eqref{ratio_mol}, which accounts for the nuclear motion, will be refered 
to as $F_{\rm nuc}$ below. Note that it was made dimensionless by inserting the 
equilibrium distance $R_{\rm eq}$ between the nuclei into the formula.

The quantum number $\nu_a$ in Eq.~\eqref{ratio_mol} determines the vibrational 
nuclear wave function of the two-center autoionizing state and fixes the precise 
value of the associated resonant transition energy. The full expression for 
$\sigma^{(2)}_{\rm mol}$ contains a coherent sum over the intermediate state and 
an incohrent sum over the final state quantum numbers, including $\nu_f$ \cite{2CPImol}. 

Before proceeding further we note that the processes of 2CPI and direct photoionization 
are generally subject to quantum interference because they lead to the same final state 
(atom $B$ merely serves as a catalyzer). 
However, for parameters where 2CPI strongly dominates, 
the interference is of minor importance and may be neglected.

\vspace{0.25cm} 

{\it Discussion---} Based on Eqs.\,\eqref{ratio_at}, \eqref{branching} and 
\eqref{ratio_mol} we can now compare the relative enhancement of 2CPI over 
direct photoionization in LiHe and NeHe dimers. 

For a system of Li and He at $R=10$\,\AA  \, considered as individual atoms, 
the ratio $\sigma_{\rm at}^{(2)}/\sigma^{(1)}_A$ in Eq.~\eqref{ratio_at} 
attains the value $\approx 4 \times 10^6$, assuming that the incident photon 
energy $\omega\approx 21.2$\,eV is resonant to the $1s\to 2p$ transition in 
helium and taking $\alpha=1$ for definiteness. In this scenario, the radiative 
width $\Gamma_r^{(2p)} = \Gamma_{r,ge}^{(2p)} \approx 1.18\times 10^{-6}$\,eV 
\cite{NIST} is much larger than the Auger width $\Gamma_a$. 
When a LiHe dimer \cite{LiHe} is considered instead, 
the single resonance splits into a multiplet of resonances, in accordance with 
the various vibrational transitions \cite{2CPImol}. On each of these resonances, 
the nuclear motion tends to reduce the enhancement but not dramatically 
($F_{\rm nuc}\sim 0.1$ for favoured transitions $\nu_i\to\nu_a\to\nu_f$). 
Averaging over the molecular orientations with respect 
to the field, when the dimers are randomly distributed, leads to a further 
reduction of the cross section by a geometrical factor of order unity.  

In the experiment \cite{2CPIexp}, Ne is ionized by synchrotron photons of energy 
$\omega \approx 23.1$\,eV which corresponds to the $1s\to3p$ transition in He 
(note that the $1s\to2p$ transition energy lies below the ionization potential of Ne). 
The largest enhancement (by a factor of $\simeq\,60$--100, 
see \cite{2CPIexp, Jabbari}) was observed when 
the intermediate state $1s3p\pi$ with $\nu_a=2$ was populated. The Auger width 
$\Gamma_a$ of this state can be estimated from the 'local' width \eqref{Gamma_a} 
which applies to a fixed value of $R$, by taking an average over the probability 
density $|\psi_a^{(\nu_a)}(R)|^2$ of the vibrational state and, accordingly, 
amounts to $\Gamma_a\approx 1$\,meV. This value agrees very well with the result 
of advanced quantum chemical calculations \cite{Jabbari}. $\Gamma_a$ 
turns out to be orders of magnitude larger than the radiative width 
$\Gamma_r^{(3p)} \approx  3.7\times 10^{-7}$\,eV \cite{NIST}, \cite{He3p}.

With these numbers, we obtain 
$\bar{\sigma}_{\rm mol}^{(2)}/\sigma^{(1)}_{\rm mol}\approx 320$,
which is by $3$--$4$ orders of magnitude smaller than in LiHe. As before, 
we have set $\alpha=1$ since the field component perpendicular to the 
molecular axis is responsible for $\pi$ state excitation. 
$\bar{\sigma}_{\rm mol}^{(2)}$ involves an average over the molecular 
orientations in the gas target and, accordingly, amounts to 2/3 of the 
cross section at $\theta_R=\frac{\pi}{2}$. Besides, the nuclear-motion 
factor was estimated as $F_{\rm nuc}\approx 1$ because the transitions 
mainly occur in a small range of internuclear distances around the equilibrium value. 

While the experimental outcome is not reproduced yet,
we already see here that the enhancement is much smaller than in LiHe: 
in a NeHe dimer $ \Gamma_a $ is huge compared to $ \Gamma_r $ and the increase 
of the branching ratio to essentially $1$ cannot compensate for the very strong 
decrease in the excitation probability caused by a very large broadening of the 
resonance due to the ICD channel. Therefore, it is this interplay between 
the photoexcitation and decay steps of 2CPI [see Eq.~\eqref{branching}] 
which is the key in explaining the counterintuitive result, 
that the relative enhancement of photoionization due to 2CPI can be much weaker 
in a relatively small NeHe-system as compared to a large LiHe dimer. 

Our description of the experiment on NeHe can be improved by noting that 
the applied synchrotron beam in \cite{2CPIexp} had a spectral width of 
$\Delta\omega = 1.7$\,meV which effectively broadens the resonance. 
At ${\rm min}$\{$\Gamma_a$, $\Delta\omega$\} $ \gg \Gamma_r$ this effect 
can be approximately taken into account by the replacement  
$\Gamma^2 \to \Gamma_a(\Gamma_a + \Delta\omega)$ in the denominator of 
Eq.~\eqref{branching}. Accordingly, it leads to a damping of 2CPI in NeHe 
by roughly a factor of $\Gamma_a/(\Gamma_a+\Delta\omega)\approx 0.37$. 
This reduces the calculated ratio 
to $\bar{ \sigma }_{\rm mol}^{(2)}/\sigma^{(1)}_{\rm mol}\simeq 118$,
which is to be compared with an approximately 60--100 fold enhancement 
observed in the experiment \cite{2CPIexp, Jabbari}.  

We point out that, from the measured data, a larger ICD width of 
$\Gamma_a\approx 2$ - $2.5 $\,meV was deduced in \cite{2CPIexp}. 
It leads to a somewhat smaller ratio of 
$\bar{\sigma}_{\rm mol}^{(2)}/\sigma^{(1)}_{\rm mol} \approx 76$ - $ 86$.

The above discussion indicates that the spectral width might have 
a large detrimental impact on the enhancement effect. However, unlike 
the ICD width $\Gamma_a$, this factor can be avoided by using coherent 
light sources with very high degree of monochromaticity. Indeed, the 
feasibility of atomic spectroscopy at $\omega \lesssim  20$\,eV and megahertz 
bandwidths ($\Delta\omega \lesssim 10^{-7}$\,eV) has been demonstrated by 
extending the frequency-comb technique into the extreme ultraviolet 
domain \cite{comb1}. Such sources can be employed for an experimental 
observation of the predicted huge enhancement of photoionization in LiHe.

\vspace{0.15cm} 

{\it Conclusion---}We have shown that the resonant enhancement of 
photoionization due to two-center dipole-dipole correlations can be 
very strongly reduced when the inter-site distance decreases, even 
though the strength of the correlations {\it per se} greatly increases. 
This counterintuitive result also applies to other resonant two-center 
phenomena, such as, for instance, interatomic photo double ionization \cite{double} 
and two-center dielectronic recombination \cite{2CPI_PRA, 2cdr}, which 
represents the inverse of 2CPI. All this shows that in order to 'extract' 
most efficiency from the resonant two-center coupling, 
the latter must not be too strong, i.e., the
interacting centers not be located too close to each other.

\section*{Acknowledgement}
This work has been funded by the Deutsche Forschungsgemeinschaft (DFG, German 
Research Foundation) under Grant No. 349581371 (MU 3149/4-1 and VO 1278/4-1). 



\begin{thebibliography}{33}

\bibitem{ICD} L. S. Cederbaum, J. Zobeley and F. Tarantelli, 
Phys. Rev. Lett. \textbf{79}, 4778 (1997).

\bibitem{ICDres} While the initial state for ICD is usually prepared 
by photoionization of an inner-valence electron, so-called resonant ICD 
occurs after photo{\it{excitation}}; see, e.g., K.~Gokhberg, A. B. Trofimov, 
T. Sommerfeld, and L.~S. Cederbaum, Europhys. Lett. {\bf 72}, 228 (2005).

\bibitem{ICDrev} For reviews on ICD, see 
U. Hergenhahn, J. Electron Spectrosc. Relat. Phenom. {\bf 184}, 78 (2011);
V. Averbukh {\it et al.}, {\it ibid.} {\bf 183}, 36 (2011);
T. Jahnke, J. Phys. B \textbf{48}, 082001 (2015).

\bibitem{clusters} S. Marburger, O. Kugeler, U. Hergenhahn, and T. M\"oller,
Phys. Rev. Lett. \textbf{90}, 203401 (2003).

\bibitem{dimers} T. Jahnke \textit{et al.}, Phys. Rev. Lett. \textbf{93}, 083002 (2004);
Y.~Morishita \textit{et al.}, {\it ibid.} \textbf{96}, 243402 (2006);
T.~Havermeier \textit{et al.}, {\it ibid.} \textbf{104}, 133401 (2010).

\bibitem{COLTRIMS}
J. Ullrich, R. Moshammer, A. Dorn, R. D\"orner, L. P. H. Schmidt, and H. Schmidt-B\"ocking, 
Rep. Prog. Phys. {\bf 66}, 1463-1545 (2003).

\bibitem{excitons}
J. Frenkel,
Phys. Rev. {\bf 37}, 17 (1931);
G. D. Scholes and G. Rumbles,
Nature Materials {\bf 5}, 683-696 (2006).

\bibitem{Rydberg} T. Amthor {\it et al.}, Phys. Rev. Lett. {\bf 98}, 023004 (2007);
C. S. E. van Ditzhuijzen, A. F. Koenderink, J. V. Hernandez, F. Robicheaux, L. D. Noordam, 
and H. B. van Linden van den Heuvell, Phys. Rev. Lett. {\bf 100}, 243201 (2008).

\bibitem{water} T. Jahnke {\it et al.}, Nature Phys. {\bf 6}, 139 (2010); 
M. Mucke {\it et al.} {\bf 6}, 143 (2010);
C. Richter {\it et al.}, Nat. Commun. {\bf 9}, 4988 (2018).

\bibitem{Dorn}
X. G. Ren, E. L. Wang, A. D. Skitnevskaya, A. B. Trofimov, K. Gokhberg, and A. Dorn,
Nat. Physics {\bf 14}, 1062 (2018)

\bibitem{polymers} S. Suhai, Phys. Rev. B {\bf 51}, 16553 (1995).

\bibitem{Forster} T. F\"orster, Ann. Phys. (Leipzig) {\bf 437}, 55 (1948);
T.~Renger, V. May and O. K\"uhn, 
Phys. Rep. {\bf 343}, 137 (2001);
E. A. Jares-Erijman and T. M. Jovin, Nature Biotechnol. {\bf 21}, 1387 (2003).

\bibitem{Gokhberg}
K. Gokhberg, P. Koloren{\v c}, A. I. Kuleff, and L. S. Cederbaum, 
Nature {\bf 505}, 661 (2014).

\bibitem{2CPI} B. Najjari, A. B. Voitkiv, and C. M\"{u}ller, 
Phys. Rev. Lett. {\bf 105} 153002 (2010).

\bibitem{2CPI_PRA}
A. B. Voitkiv and B. Najjari, Phys. Rev. A \textbf{82}, 052708 (2010).

\bibitem{Perina} J. Pe\v{r}ina, A. Luk\v{s}, V. Pe\v{r}inov{\'a}, 
and W. Leo{\'n}ski, Phys. Rev. A {\bf 83}, 053416 (2011); 
V. Pe\v{r}inov{\'a}, A. Luk\v{s}, J.~K\v{r}epelka, and 
J. Pe\v{r}ina, {\it ibid.} \textbf{90} 033428 (2014).

\bibitem{2CPIcoll} A. B. Voitkiv, C. M\"uller, S. F. Zhang, and X. Ma,
New J. Phys. 21, 103010 (2019)

\bibitem{2CPImol} F. Gr\"ull, A. B. Voitkiv, and C. M\"uller, 
submitted (2020) (preprint available on arXiv:2004.02459).

\bibitem{2CPIexp} F. Trinter \textit{et al.}, Phys. Rev. Lett. \textbf{111} 233004 (2013).

\bibitem{Jabbari} G. Jabbari, S. Klaiman, Y.-C. Chiang, F. Trinter, 
T.~Jahnke, and K. Gokhberg, J. Chem. Phys. {\bf 140}, 224305 (2014).

\bibitem{Mhamdi2018} A. Mhamdi \textit{et al.}, Phys. Rev. A {\bf 97}, 053407 (2018).

\bibitem{Sisourat2010}
N. Sisourat, H. Sann, N. V. Kryzhevoi, P. Koloren{\v c}, T.~Havermeier, 
F. Sturm, T. Jahnke, H.-K. Kim, R.~D\"orner and L. S. Cederbaum,
Phys. Rev. Lett. {\bf 105}, 173401 (2010).

\bibitem{Hergenhahn} A. Hans, P. Schmidt, C. Ozga, C. Richter, H. Otto, 
X.~Holzapfel, G. Hartmann, A. Ehresmann, U. Hergenhahn and A. Knie, 
J. Phys. Chem. Lett. \textbf{10}, 1078 (2019).

\bibitem{He-dimer} Signatures of the nuclear dynamics have also been identified 
in ICD in He dimers; see
T. Havermeier {\it et al.}, Phys. Rev. Lett. {\bf 104}, 133401 (2010);
N. Sisourat, N.~V. Kryzhevoi, P. Koloren{\v c}, S. Scheit, and L. S. Cederbaum, 
Phys. Rev. A {\bf 82}, 053401 (2010);
A. Mhamdi, J.~Rist, T.~Havermeier, R. D\"orner, T. Jahnke, and P. V. Demekhin,
Phys. Rev. A {\bf 101}, 023404 (2020).

\bibitem{Gamma} Since we treat 2CPI in the second order of perturbation theory, 
the width $\Gamma$ of the excited state in atom $B$ must be added by 'hand'. 
When the process is described, instead, within the theory of Fano resonances 
\cite{2CPI_PRA}, the excited state automatically receives a non-zero width.

\bibitem{eff_pol} Note that Eq.~\eqref{CS} contains an effective polarization vector 
${\bf n}_{\rm eff} = {\bf e}_z - 3\cos\theta_R\,{\bf e}_R$ which depends on the 
field polarization and the relative interatomic orientation. It may prove useful 
for the interpretation of angular emission patterns from 2CPI \cite{Mhamdi2018}.

\bibitem{e-impact} F. Gr\"ull, A. B. Voitkiv and C. M\"uller, 
Phys. Rev. A {\bf 100}, 032702 (2019).

\bibitem{exc} B. H. Bransden and C. J. Joachain, {\it Physics of Atoms and Molecules} 
(Longman Group, Harlow, 1990).

\bibitem{NIST} Atomic spectra database of the National Institute of Standards 
and Technology (NIST), available at https://www.nist.gov/pml/atomic-spectra-database

\bibitem{LiHe} B. Friedrich, Physics {\bf 6}, 42 (2013).

\bibitem{He3p} The $1s3p$ state in He decays radiatively to the $1s^2$ ($\Gamma_{r,ge}^{(3p)}$) 
and $1s2s$ states. However, the contribution to $\Gamma_{r}^{(3p)}$ from the latter is smaller 
than $\Gamma_{r,ge}^{(3p)}$ by a factor of about 50.

\bibitem{comb1}
A. Cing\"oz, D. C. Yost, T. K. Allison, A. Ruehl, M. E. Fermann, I. Hartl, and J. Ye, 
Nature {\bf 482}, 68 (2012).

\bibitem{double} A. C. LaForge, M. Shcherbinin, F. Stienkemeier, R.~Richter, 
R. Moshammer, T. Pfeifer and M. Mudrich, Nature Phys. {\bf 15}, 247 (2019);
A. Eckey, A. B. Voitkiv and C. M\"uller, J. Phys. B {\bf 53}, 055001 (2020).

\bibitem{2cdr} C. M\"uller, A. B. Voitkiv, J. R. Crespo Lopez-Urrutia, 
Z.~Harman, Phys. Rev. Lett. {\bf 104}, 233202 (2010).

\end{thebibliography}
\end{document}